\documentclass[preprint,preprintnumbers,amsmath,amssymb,showkeys]{revtex4}

\usepackage{amsmath}
\usepackage{upgreek}

\usepackage{dcolumn}
\usepackage{bm}
\usepackage{graphicx}
\usepackage{epstopdf}
\def\bond#1{\relax
 \hbox \bgroup
      \kern 0.1em
      \setbox0=\hbox{X}%
      \vbox to \ht0 \bgroup
         \vss
         \ifcase#1
         \or
            \hrule width .7em depth 0pt height 0.05em
         \or
            \hrule width .7em depth 0pt height 0.05em
            \vskip 0.2em
            \hrule width .7em depth 0pt height 0.05em
         \fi
         \vss
         \egroup
      \kern 0.1em
      \egroup
   }

\newcommand{\nl}{\nonumber \\}
\newcommand{\ie}{{\rm i.e.}}

\newcommand{\be}{\begin{equation}}
\newcommand{\ee}{\end{equation}}
\newcommand{\bea}{\begin{eqnarray}}
\newcommand{\eea}{\end{eqnarray}}
\newcommand{\Fig}[1]{Fig.\,\ref{#1}}
\newcommand{\Eq}[1]{Eq.\,(\ref{#1})}
\newcommand{\Eqs}[1]{Eqs.\,(\ref{#1})}
\newcommand{\Sch}{Schr\"{o}dinger\ }

\newcommand{\Matr}{\textbf {M}}
\newcommand{\MatrMx}{\textbf {M}_{\rm X}}
\newcommand{\MatrA}{\textbf {A}}
\newcommand{\MatrB}{\textbf {B}}
\newcommand{\MatrC}{\textbf {C}}
\newcommand{\MatrD}{\textbf {D}}
  \def\bP{\mbox{\boldmath $P$}}
  \def\bR{\mbox{\boldmath $R$}}
  \def\bQ{\mbox{\boldmath $Q$}}
  \def\bfe{\mbox{\boldmath $e$}}
  \def\bfX{\mbox{\boldmath $X$}}
  \def\bfJ{\mbox{\boldmath $J$}}
  \def\bU{\textbf{U}}
  \def\bM{\textbf{M}}
  \def\bG{\textbf{G}}
\def\cm{cm$^{-1}\,$}

\begin{document}

\preprint{}

\title{A Concise Method for Kinetic Energy Quantisation}

\author{Yonggang Yang}
 \email{ygyang@chemie.fu-berlin.de}
\author{Oliver K\"{u}hn}%
\altaffiliation[New Address:]{$\,$Institut f\"ur Physik, Universit\"at Rostock, Universit\"atsplatz 3, D-18051 Rostock, Germany, Tel. +493814986950, Fax: +493814986942} \email{oliver.kuehn@uni-rostock.de}

\affiliation{
Institut f\"{u}r Chemie und Biochemie, Freie Universit\"{a}t Berlin, Takustr. 3, D-14195 Berlin, Germany
}%

\date{\today}

\begin{abstract}
We present a straightforward method for obtaining exact classical and quantum molecular Hamiltonians in terms of arbitrary coordinates. As compared to other approaches the resulting expression are rather compact, the physical meaning of each quantity is quite transparent, and in some cases  the calculation effort will be greatly reduced. We also investigate systems with constrains to find the suggested method to be applicable in contrast to  most conventional approaches to kinetic energy operators which cannot  directly be applied to constrained systems. Two examples are discussed in detail, the monohydrated hydroxide anion and the protonated ammonia dimer.
\end{abstract}

\maketitle
%
\section{\label{sec:intro}Introduction}
%
Theoretical modeling of moelcular reaction dynamics requires to have at hand a Hamiltonian. For few-atom systems this may involve defining the potential energy surfaces (PES) explicitly in terms of properly chosen coordinates such as bond length and angles or distance vectors. In the latter case, for instance, it has been shown that PES in full-dimensionality including proper account of exchange symmetry can be generated for systems like the five-atomic H$_{3}$O$_{2}^-$ \cite{huang04:5042} or the seven-atomic H$_{5}$O$_{2}^+$ \cite{huang05:044308}. With increasing dimensionality this approach is no longer feasible due to the exponential scaling of the computational effort for explicit calculation of the PES on a numerical grid. Here, reaction surface concepts have been proven to be a versatile compromise, mapping out that PES region which is relevant for the considered reaction by combining suitable large amplitude reaction coordinates with orthogonal harmonic motions\cite{miller80:99,carrington84:3942,shida89:4061,tew01:1958,giese05:054315}. 

Besides the PES the kinetic energy operator (KEO) needs to be known in terms of the chosen coordinate system. In principle, kinetic energy quantisation had been discussed by Podolsky in terms of  intricate tensor analysis right after the foundation of the \Sch equation  had been laid \cite{podolsky28:812}. Later studies are to some extent based on the same mathematical techniques \cite{mecke36:405,bunker79}. 
{\bf Approximate KEOs using normal mode coordinates are widely adopted which
go back to contributions by  Eckart \cite{eckart35:552}, Wilson \cite{wilson55}, and Watson \cite{watson68:479}. 
In particular, the Eckart equations \cite{eckart35:552} enable one to determine an orientation of molecule fixed axes suitable  for using normal modes. The details of the derivation of a rovibrational Hamiltonian can be found in Ref. \cite
{wilson55}. A significant simplification of this rovibrational Hamiltonian in terms of normal coordinates has been achieved by Watson
\cite{watson68:479,watson70:465}. In fact the Watson Hamiltonian served as a starting point for many later investigations and is implemented, e.g., in the MULTIMODE program of Bowman, Carter, and coworkers \cite{bowman03:533}.
In principle normal mode coordinates provide a very compact representation of the PES which is taylored, however, to the stationary point for which they have been defined. For large amplitude motions away from this stationary point the above mentioned reaction surface approach or the use of general curvilinear  coordinates are more suitable. A detailed discussion of advantages and disadvantages of various
coordinates for rovibrational Hamiltonians has been presented, e.g.,  in Ref.
\cite{bramley91:1183}.}

In recent years Gatti and coworkers have developed a general scheme for obtaining a KEO using vector parameterization, \ie, the non-trivial coordinates are expressed with $N-1$ real vectors for system of $N$ atoms \cite{gatti01:8275,gatti03:507,iung06:130}. This scheme is quite successful and convenient for a full dimensional description \cite{vendrell07:184302}. However, often one needs a reduced dimensional description leading to a problem with imposed constraints \cite{nauts87:164,gatti97:403}. Generally speaking in this case the coordinates can not be expressed as components of real vectors. Finally, we should point out that a numerical alternative to the analytical determination of the KEO has been suggested in Ref. \cite{lauvergnat02:8560}.

In the present work we introduce a  simple physical method of kinetic energy quantisation which directly quantises the classical Lagrange/Hamilton functions. The formal theory is presented in Section \ref{sec:general}. Here it is also shown that our method can be directly applied to constrained systems. In Section \ref{sec:app} the approach is applied to two molecular systems, that is, H$_{3}$O$_{2}^-$ which is treated in full dimensionality and N$_{2}$H$_{7}^+$ for which certain constraints are applied. 
%
\section{General Theory for Obtaining a KEO}
\label{sec:general}
%
\subsection{Formal Development}\label{fulldim}
In the following we focus  on the most common cases where the kinetic energy includes purely quadratic terms and the potential energy does not depend on the velocity, although a generalisation would be straightforward. Further we adopt the laboratory reference frame (LRF). The classical Lagrange/Hamiltonian in terms of arbitrary variables can be written as
\bea\label{lagrange}
L(\bQ,\dot {\bQ},t)
&=&T(\dot{\bQ})-V(\bQ)
=\frac{1}{2} \dot {\bQ} ^\dagger \Matr \dot {\bQ}-V(\bQ)  \nl
H(\bQ,\bP)
&=&T(\bP)+V(\bQ)
=\frac{1}{2}\bP ^\dagger \Matr^{-1} \bP+V(\bQ),
\eea
where $\bQ$ and $\bP$ are single column vectors of the generalised coordinates and corresponding conjugate momenta, respectively. The generalised momentum vector is defined as $\bP=\partial L/\partial \dot {\bQ}=\Matr \dot {\bQ}$. 
The generalised mass matrix, $\Matr$, in general is a function of the coordinates. We assume that it is defined such as to be Hermitian and ``$\dagger$'' means Hermitian conjugate which is equivalent to the transpose in classical mechanics.  In order to obtain the quantum mechanical Hamiltonian operator, the major task is to derive the quantum KEO in coordinate representation. 

Since the choice of coordinates is quite arbitrary in \Eq{lagrange}, we start with Cartesian ones \{$\bfX$,$\bP_X$\}; 
in the following all the subscripts ``$_X$'' are associated with Cartesian coordinates. The operator for each component of the momentum $(\hat{\bP}_X^\dagger)_j=(\hat{\bP}_X)_j =-i \hbar \partial/\partial X_{j}$ is Hermitian.  For non-Cartesian coordinates we define the momentum operator in the same way:
\be\label{1pquam}
\hat{\bP}=-i \hbar \frac{\partial}{\partial \bQ}. 
\ee
To obtain the quantum KEO in terms of non-Cartesian coordinates, we must exploit the coordinate transform relations, $\bQ=\bQ(\bfX)$. The following relations can be derived easily
\bea
\label{eq:coortrans}
&&\dot {\bQ}=\frac{\partial \bQ}{\partial \bfX}
\dot{\bfX} \nl
&&\hat{\bP}=\left(\frac{\partial \bfX}{\partial \bQ}\right) ^{\dagger}  
\hat{\bP}_X \nl
&&\frac{\partial \bQ}{\partial \bfX}
=\left(\frac{\partial \bfX}{\partial \bQ}\right) ^{-1},
\eea
where the elements of the transformation matrices are defined as $\left(\partial \bQ/\partial \bfX\right)_{ij}=\partial Q_i/\partial X_j$ and 
$\left(\partial \bfX/\partial \bQ \right)_{ij}=\partial X_i/\partial Q_{j}$.
The quantum KEO in terms of non-Cartesian coordinates can be obtained by the simple coordinate transform from Cartesian coordinates
\bea\label{ketrans}
\hat{T}&=&\frac{1}{2}\hat{\bP}_X ^\dagger \MatrMx^{-1} \hat{\bP}_X \nl
&=&\frac{1}{2}\hat{\bP} ^\dagger \frac{\partial \bQ}{\partial \bfX} \MatrMx^{-1}
\left(\frac{\partial \bQ}{\partial \bfX}\right) ^{\dagger}  \hat{\bP },
\eea
where $\MatrMx$ is the diagonal Cartesian mass matrix consisting of real mass associated with each Cartesian coordinate.

The momentum operator vector $\hat{\bP}$ can be replaced by the derivative operator $\hat{\bP}=-i \hbar \partial/\partial \bQ$. 
However, the expression for the Hermitian conjugate of momentum operator (HCMO) vector
$\hat{\bP} ^\dagger$ needs to be established yet. Since one has the relation between $\hat{\bP}$ and the Cartesian one $\hat{\bP}_X$
it is not difficult to directly apply the definition of Hermitian conjugation to Eq. (\ref{eq:coortrans})
\bea\label{pdagger}
\left(\hat{\bP} ^\dagger\right)_i
&=&\left(\left(\left(\frac{\partial \bfX}{\partial \bQ}\right) ^\dagger
 \hat{\bP}_X\right)^\dagger\right)_i 
=\left(\hat{\bP}_X ^\dagger \frac{\partial
\bfX}{\partial \bQ}\right)_i \nl
&=&\sum_{j}
\left(\hat{\bP}_X \right)_j \left(\frac{\partial
\bfX}{\partial \bQ}\right)_{ji} \nl
&=&\sum_{j}
\left(\frac{\partial \bfX}{\partial \bQ}\right)_{ji}
\left(\hat{\bP}_X \right)_j
+\sum_{j}
\left[\left(\hat{\bP}_X \right)_j,\left(\frac{\partial
\bfX}{\partial \bQ}\right)_{ji}\right] \nl
&=&\left(\hat{\bP}\right)_i+\sum_{j}
\left[\left(\hat{\bP}_X \right)_j,\left(\frac{\partial
\bfX}{\partial \bQ}\right)_{ji}\right],
\eea  
where the notation $[\hat{A},\hat{B}]$ means the commutator of two operators. Eq. (\ref{pdagger}) shows the non-Hermiticity of generalised momenta associated with non-Cartesian coordinates. Additional terms appear due to the non-commutability of the momenta and the transformation matrix. Using the basic commutator $[\hat{X}_j,(\hat{P}_X)_k]=i \hbar \delta _{jk}$, Eq. (\ref{pdagger}) can be simplified to (see also Ref. \cite{lyi96:1251})
\begin{equation}
\label{pdgmore}
(\hat{\bP} ^\dagger)_k=(\hat{\bP})_k-i \hbar
\sum_j \left(\frac{\partial}{\partial {X_j}} \frac{\partial
{X_j}}{\partial {Q}_k}\right)^\circ.
\end{equation}
Here, the superscript $^\circ$ means a differential operator inside the bracket can not operate on
functions outside, in other words,  the result is just normal function of coordinates.

\Eq{pdgmore} clearly shows the relation between a momentum operator and its Hermitian conjugate. Apparently, 
any set of functions of coordinates, $\{f_j(Q_j)\}$, can be multiplied from the left to the derivative operator in \Eq{1pquam} yielding different schemes of momenta quantisation, \ie, $(\hat{\bP})_j =-i \hbar \cdot f_j(Q_j) \cdot \partial/\partial Q_{j}$ constitutes also an acceptable scheme. 
Since these momentum operators are in general non-Hermitian, we cannot simply set $f_j(Q_j)=1$.
For different schemes of momenta quantisation, the HCMOs will vary correspondingly to keep the KEO invariant. But, this enables us to optimise momenta quantisation schemes. In other words with proper functions one can obtain desired forms of momentum operators, e.g., symmetric forms.

With \Eq{ketrans} and \Eq{pdagger} we get the general scheme of constructing
KEOs in terms of arbitrary coordinates. 
However, the final structure seems to be complicated and actually it can be
simplified. To this end we compare \Eq{ketrans} and \Eq{lagrange} and take into account the invariance of the classical kinetic energy 
\bea
T&=&\frac{1}{2} \dot
{\bQ} ^\dagger \Matr \dot {\bQ} \nl
&=&\frac{1}{2}
\dot {\bfX} ^\dagger \MatrMx \dot {\bfX} \nl
&=&\frac{1}{2}
\dot {\bQ} ^\dagger 
\left(\frac{\partial \bfX}{\partial \bQ}\right)^\dagger \MatrMx 
\frac{\partial \bfX}{\partial \bQ} \dot {\bQ}. 
\eea
Since $\dot {\bQ}$ can be any vector we must have the relation $\Matr=\left(\partial \bfX/\partial \bQ\right)^\dagger \MatrMx \partial \bfX/\partial \bQ$.
Calculating the inverse of both sides we immediately see that the complicated central part 
$(\partial \bQ/\partial \bfX) \MatrMx^{-1}\left(\partial \bQ/\partial \bfX\right) ^\dagger$  in \Eq{ketrans} is exactly $\Matr^{-1}$ in \Eq{lagrange}. 
The final KEO is therefore simplified as
\be
\hat{T}=\frac{1}{2}\hat{\bP} ^\dagger \Matr^{-1} \hat{\bP}.
\ee
This means we can directly exploit the result of \Eq{lagrange} and quantise the generalised momenta without the knowledge of the Cartesian kinetic energy. 
All we need are the generalised mass matrix $\Matr$ and the definition of coordinates which both appear in \Eq{lagrange}.

Therefore we can identify a concise and physically transparent scheme to construct the KEO:
\begin{enumerate}
\item[1.]Get the classical kinetic energy and make sure the mass matrix is symmetric, so that one has the same structure as \Eq{lagrange}, $T(\dot{\bQ})
=\frac{1}{2} \dot {\bQ} ^\dagger \Matr \dot {\bQ}$.
\item[2.]Exploit \Eq{pdagger} or \Eq{pdgmore} to express HCMOs $\hat{\bP} ^\dagger$ in terms of $\hat{\bP}$ and some functions of $\bQ$.
\item[3.]Calculate the inverse matrix of $\Matr$ hence the formal quantum KEO reads $\hat{T}=\frac{1}{2}\hat{\bP} ^\dagger \Matr^{-1} \hat{\bP}$.
\item[4.]Replace $\hat{\bP}$ by $-i \hbar \partial/\partial \bQ$ for obtaining the coordinate representation.
\end{enumerate}
The first step is quite familiar and one can choose arbitrary coordinates to get the classical kinetic energy.
Actually we will introduce a useful partition method in Section \ref{partitionke} which will significantly simplify this issue. 
The remaining three steps are quite straightforward to follow. Apparently, the major effort is in the second step, i.e. determining the HCMOs according to \Eq{pdgmore}.
However, for the familiar spherical coordinates the HCMOs are well-known which leads to a great simplification as shown in Section \ref{hcmo}.

Apart from the KEO the volume element of integration is also of importance. Since we start from Cartesian coordinates and all that we have done is a coordinate transformation, the Euclidean normalisation remains correct: 
\begin{equation}
d \tau = d \tau _X = \prod_i d X_i = |\mathrm{Det}\left(\frac{\partial \bfX}{\partial \bQ}\right)| \prod_i d Q_{i}.
\end{equation}
\subsection{\label{reduceddim}Systems with Constraints}
Reduced dimensional descriptions are always necessary for large systems. 
In the following we will give the general description for systems with constraints. 
Considering a system with some active coordinates $\bQ_1$ and some frozen coordinates $\bQ_0$, the full dimensional coordinates and corresponding conjugate momenta read
\begin{equation}
\begin{array}{ccccc}
\bQ=\left(
\begin{array}{c}
\bQ_1\\
\bQ_0
\end{array}
\right) &&&&
\bP=\left(
\begin{array}{c}
\bP_1\\
\bP_0
\end{array}
\right)
\end{array}.
\end{equation}
The constraining conditions are given by $\dot{\bQ}_0=0$. Thus we can obtain the constrained classical kinetic energy
\bea\label{lagrred}
T&=&\frac{1}{2} \dot {\bQ} ^\dagger \Matr \dot {\bQ} \nl
&=&\frac{1}{2}
\left(
\begin{array}{cc}
\dot {\bQ}_1^\dagger & \dot {\bQ}_0^\dagger
\end{array}
\right)
\left(
\begin{array}{cc}
\Matr_{11} & \Matr_{10}\\
\Matr_{01} & \Matr_{00}
\end{array}
\right)
\left(
\begin{array}{c}
\dot {\bQ}_1\\
\dot {\bQ}_0
\end{array}
\right) \nl
&=&\frac{1}{2} \dot {\bQ}_1 ^\dagger \Matr_{11} \dot {\bQ}_1,
\eea
where  $\Matr_{ij}$ are corresponding sub-matrices. 

To get the quantum KEO we have to rewrite the constraint conditions in terms of momenta. 
According to the definition of momenta it is not difficult to find the following relation
\bea
\left(
\begin{array}{c}
\dot {\bQ}_1 \\
\dot {\bQ}_0
\end{array}
\right)&=&
\left(
\begin{array}{cc}
\Matr_{11} & \Matr_{10}\\
\Matr_{01} & \Matr_{00}
\end{array}
\right) ^{-1}
\left(
\begin{array}{c}
\bP_1\\
\bP_0
\end{array}
\right) \nl
&=&
\left(
\begin{array}{cc}
\MatrA & \MatrB\\
\MatrC & \MatrD
\end{array}
\right)
\left(
\begin{array}{c}
\bP_1\\
\bP_0
\end{array}
\right),
\eea
where $\left(
\begin{array}{cc}
\MatrA & \MatrB\\
\MatrC & \MatrD
\end{array}
\right)$ is the inverse of matrix $\Matr$. 
Thus one can rewrite the constraint conditions as $\MatrC \bP_1+\MatrD \bP_0=0$ or $\bP_0=-\MatrD ^{-1}\MatrC \bP_1$. 
After quantization we get the constraint relation for the corresponding quantum operators, \ie, $\hat{\bP}_0=-\MatrD ^{-1}\MatrC \hat{\bP}_1$. 
Based on this observation we can obtain the quantum KEO
\bea\label{kereduced}
\hat{T}&=&\frac{1}{2} {\hat{\bP}} ^\dagger \Matr^{-1} {\hat{\bP}} \nl
&=&\frac{1}{2}
\left(
\begin{array}{cc}
\hat{\bP}_1^\dagger & \hat{\bP}_0^\dagger
\end{array}
\right)
\left(
\begin{array}{cc}
\Matr_{11} & \Matr_{10}\\
\Matr_{01} & \Matr_{00}
\end{array}
\right) ^{-1}
\left(
\begin{array}{c}
\hat{\bP}_1\\
\hat{\bP}_0
\end{array}
\right) \nl
&=&\frac{1}{2}
\left(
\begin{array}{cc}
\hat{\bP}_1^\dagger & (-\MatrD ^{-1}\MatrC \hat{\bP}_1)^\dagger
\end{array}
\right)
\left(
\begin{array}{cc}
\MatrA & \MatrB\\
\MatrC & \MatrD
\end{array}
\right)
\left(
\begin{array}{c}
\hat{\bP}_1\\
-\MatrD ^{-1}\MatrC \hat{\bP}_1
\end{array}
\right) \nl
&=&\frac{1}{2} \hat{\bP}_1 ^\dagger \left(\MatrA-\MatrB\MatrD^{-1}\MatrC \right) 
\hat{\bP}_1\nl
&=&\frac{1}{2} \hat{\bP}_1 ^\dagger \Matr_{11}^{-1} \hat{\bP}_1.
\eea

The final result is quite compact.
Comparing \Eq{lagrred} with \Eq{kereduced} we observe that the \emph{same procedure} mentioned in the full dimensional case can be followed provided we only consider the active coordinates and completely ignore the frozen ones from the very beginning when we generate the classical Lagrangian.
This is an attractive point since when we need to deal with constrained systems the present scheme only requires the classical Lagrangian which can be obtained by traditional methods.
\subsection{\label{partitionke} A General Method for Partitioning the Classical Kinetic Energy }
So far we have given the general theory of kinetic energy quantization starting from the classical Lagrangian.
In the following we will introduce a method for obtaining the classical kinetic energy with a partitioning technique which will greatly simplify the problem in most cases. In general it is quite convenient to divide a large system into small subsystems especially when a subsystems has certain symmetry.
If we arbitrarily divide a system into $N$ parts the kinetic energy is a sum of the $N$ subsystems. According to the K\"{o}nig theorem, 
 \be
T=\sum_{i=1}^N T_i^0=T_{(C)}^0+\sum_{i=1}^N T_i^{(C)},
\ee
where $T_A^B$ is the kinetic energy of the part $A$ with respect to the reference frame defined by $B$. Here $0$ is the laboratory reference frame and $(C)$ is the centre of mass reference frame.

In the special case when $N=2$ the following relation can be obtained
 \be\label{part-KE}
\sum_{i=1}^N T_i^{(C)}=T_1^{(C)}+T_2^{(C)}=T_1^{(C_1)}+T_2^{(C_2)}+T_{(C_1)}^{(C_2)},
\ee
where $(C_i)$ is the centre of mass of the $i$th part and $T_{(C_1)}^{(C_2)}$ is the kinetic energy of the centre of mass of the first part with respect to the centre of mass of the second part.  If the two parts are both mass points \Eq{part-KE} can be simplified as follows
\be
T_1^{(C)}+T_2^{(C)}=T_{(C_1)}^{(C_2)}=T_{(C_2)}^{(C_1)},
\ee
which is a familiar result from two-body mechanics. By exploiting \Eq{part-KE} repeatedly we can easily express the total kinetic energy in terms of kinetic energies of subsystems which are much easier to obtain as will be seen in the following.
\subsection{Hermitian Conjugates of Momentum Operators}\label{hcmo}
As mentioned in Section \ref{fulldim}, the most tedious task for generating a KEO is to derive the expressions for the HCMOs. As an example let us consider a system described by the 3D spherical coordinates \{$R$, $\theta$, $\varphi$\}. The coordinate transformation between 3D spherical coordinates and 3D Cartesian ones \{$x$, $y$, $z$\} is defined as
\bea
x&=&R \sin \theta \cos\varphi \nl
y&=&R \sin \theta \sin\varphi \nl
z&=&R \cos \theta.
\eea
Here, we only give the corresponding HCMOs leading to the well-known KEO (for details, see Appendix \ref{sec:appd})
\bea
\hat{P}_{R}^{\dagger}&=&\hat{P}_{R}-\frac{2i \hbar}{R}
=-i\hbar \frac{1}{R^2} \frac{\partial}{\partial R} R^2 \nl
\hat{P}_{\theta}^{\dagger}&=&\hat{P}_{\theta}-i \hbar \cot \theta
=-i\hbar\frac{1}{\sin\theta} \frac{\partial}{\partial \theta} \sin\theta \nl
\hat{P}_{\varphi}^{\dagger}&=&\hat{P}_{\varphi}
=-{i\hbar}\frac{\partial }{\partial \varphi} \nl
\hat{T}&=&
-\frac{\hbar^2}{2m} \frac{1}{R^2} \frac{\partial }{\partial R} R^2 \frac{\partial }{\partial R}
-\frac{\hbar^2}{2mR^2}\frac{1}{\sin\theta} \frac{\partial}{\partial \theta} \sin\theta 
\frac{\partial }{\partial \theta}
-\frac{\hbar^2}{2mR^2 \sin^2\theta} \frac{\partial^2}{\partial \varphi^2}.
\eea

So far our considerations have been concerned with the LRF. However, in order to describe a molecule in terms of its natural motions, e.g., bond lengths and bond angles, it is usually more convenient to use one or more molecular reference frames (MRFs). 
In the following we will discuss this point and derive results for HCMOs associated with MRF spherical coordinates. 
Consider the LRF and a MRF defined by sets of unit vectors $\{\bfe_{x}, \bfe_{y}, \bfe_{z}\}$ and $\{\bfe_{x'}, \bfe_{y'}, \bfe_{z'}\}$, respectively. 
The relation between the LRF and the MRF is just an orthogonal transformation characterised by the three Euler angles \{$\vartheta$, $\phi$, $\chi$\}
\be\label{LRF2MRF}
\bfe_{\alpha'} = \bU_z(\phi) \bU_y(\vartheta) \bU_z(\chi) \bfe_{\alpha},
\ee
where $ \alpha=x,y,z$ and $\bU_{\alpha}$ is a rotation around $\bfe_{\alpha}$. (The expression for the rotational transformation matrix $\bU_{\alpha}$ and more details on the following derivations can be found in the Appendix \ref{sec:appd}.)

Now let us consider a vector $\bR_j$ characterised by three spherical coordinates \{$R_j$, $\theta_j$, $\varphi_j$\} in the MRF. 
According to Appendix \ref{sec:appd}, \Eq{aacoortrans1}, we can express $\bR_j$ as
\bea\label{eqmf1}
\bR_j=R_j \bU_z(\phi) \bU_y(\vartheta) \bU_z(\chi) \bU_z(\varphi_j) \bU_y(\theta_j) \bfe_z.
\eea
This is a vector equation and we can obtain the Cartesian coordinates of $\bR_j$ in LRF by projecting the equation onto each LRF axis. 
Based on \Eq{eqmf1} we can exploit \Eq{pdgmore} to derive the expressions for HCMOs associated with MRF coordinates.
The final results are quite concise and they actually have the same form as those associated with LRF spherical coordinates.
That is to say, that for the momentum operators associated with $\bR_j$ ($R_j$, $\theta_j$, $\varphi_j$) the following relations are valid
\bea
\hat{P}_{R_j}^{\dagger}&=&\hat{P}_{R_j}-\frac{2i \hbar}{R_j}
=-i\hbar \frac{1}{R_j^2} \frac{\partial}{\partial R_j} R_j^2 \nl
\hat{P}_{\theta_j}^{\dagger}&=&\hat{P}_{\theta_j}-i \hbar \cot \theta_j
=-i\hbar\frac{1}{\sin\theta_j} \frac{\partial}{\partial \theta_j} \sin\theta_j \nl
\hat{P}_{\varphi_j}^{\dagger}&=&\hat{P}_{\varphi_j}
=-{i\hbar}\frac{\partial }{\partial \varphi_j}.
\eea
In Appendix \ref{sec:appd} we also confirm that the momentum operators associated with MRF Cartesian coordinates are Hermitian (cf. \Eq{eq:cartHerm}). This result is quite important as it shows that one does not need to take the effort to derive the expressions for HCMOs provided that one uses spherical coordinates, Cartesian coordinates or combinations of both, no matter whether they are defined in the LRF or MRFs.
\subsection{Angular Momentum and Rotation}\label{angular}
For a system with $N$ atoms we have $3N$ degrees of freedom (DOFs).
They are normally divided into three translational DOFs, three rotational and $3N-6$ vibrational ones.
The three translational DOFs can be separated while the remaining $3N-3$ DOFs are coupled.
In general rotational excitation energies are quite small as compared with vibrational ones suggesting an approximate separation.
For the following discussion it will be convenient  to express the KEO as sum of rotational (three angles) and vibrational ($3N-6$ other coordinates) parts as well as their coupling. Specifically, we exploit the 3D vector $\bQ_{\rm rot}$ defined as
\be
\label{eq:qrot}
\bQ_{\rm rot}^\dagger=(
\begin{array}{ccc}
\vartheta & \phi & \chi
\end{array}
) 
\ee
containing the three Euler angles which connect the LRF and MRF according to \Eq{LRF2MRF}. 
The kinetic energy and total angular momentum $\bfJ$ are defined as
\bea
2T&=&\sum_{i=1}^N m_i \dot {\bR}_i^\dagger \dot {\bR_i} \nl
\bfJ&=&\sum_{i=1}^N m_i \bR_i \times \dot {\bR_i} \, .
\eea
With $\dot {\bQ}_{\rm rot}$ being the  angular velocity of MRF the velocities can be re-expressed as 
\be\label{velocityRi}
\dot {\bR_i} = \dot {\bQ}_{\rm rot} \times \bR_i + \dot {\bR_i'},
\ee
where $\dot {\bR_i'}$ is the velocity of $\bR_i$ measured in the MRF.
With the help of \Eq{velocityRi} and some vector algebra we can rewrite the kinetic energy and angular momentum as
\bea\label{cKEandJ}
2T&=&\sum_{i=1}^N m_i\left(
R_i^2 \dot {\bQ}_{\rm rot}^\dagger\dot {\bQ}_{\rm rot} - (\dot{\bQ}_{\rm rot}^\dagger\bR_i)^2 + \dot {\bR_i'}^\dagger \dot {\bR_i'}
+ 2\dot{\bQ}_{\rm rot}^\dagger(\bR_i\times\dot {\bR_i'}) \right) \nl
\bfJ&=&\sum_{i=1}^N m_i\left(
R_i^2 \dot {\bQ}_{\rm rot} - (\dot{\bQ}_{\rm rot}^\dagger\bR_i)\bR_i  + (\bR_i\times\dot {\bR_i'})
\right) .
\eea
Using \Eqs{cKEandJ} it is straightforward to derive the following relation
\be
\bfJ=\frac{\partial T}{\partial \dot {\bQ}_{\rm rot}},
\ee
which is exactly the definition of the generalised momentum vector $\bP_{\rm rot}$ associated with the three Euler angles.

Now we can draw the following important conclusion. If the set of coordinates contains the three Euler angles \{$\vartheta$, $\phi$, $\chi$\}, the total angular momentum vector is just the generalised momentum vector associated with the three Euler angles
\bea\label{J-P1}
J_{\vartheta}&=&P_{\vartheta}=\frac{\partial T}{\partial \dot {\vartheta}} \nl
J_{\phi}&=&P_{\phi}=\frac{\partial T}{\partial \dot {\phi}} \nl
J_{\chi}&=&P_{\chi}=\frac{\partial T}{\partial \dot {\chi}} \nl
\bfJ&=&\bfe_{\dot {\vartheta}}P_{\vartheta}+\bfe_{\dot {\phi}}P_{\phi}+\bfe_{\dot {\chi}}P_{\chi}.
\eea
However, since the direction of angular velocities \{$\bfe_{\dot {\vartheta}}$, $\bfe_{\dot {\phi}}$, $\bfe_{\dot {\chi}}$\} are complicated, it is better to transform the expressions to the Cartesian components in the LRF \cite{wilson55}, \ie
\bea\label{J-P}
J_{x}&=&\sin\chi P_{\vartheta} - \csc\vartheta \cos\chi P_{\phi} + \cot\vartheta \cos\chi P_{\chi} \nl
J_{y}&=&\cos\chi P_{\vartheta} + \csc\vartheta \sin\chi P_{\phi} - \cot\vartheta \sin\chi P_{\chi} \nl
J_{z}&=&P_{\chi}.
\eea
Using \Eq{J-P} it is straightforward to express the KEO in terms of total angular momentum and generalised momenta associated with vibrational DOFs.
In other words, starting from \Eq{cKEandJ} the KEO can be readily expressed as contributions of rotational part, vibrational part and their coupling. 

If the focus is on the vibrational spectrum, there are several arguments for neglecting rotational excitations. There is usually a clear time scale separation, that is, rotational excitation energies are much smaller than vibrational ones. If we are not aiming at high resolution spectroscopy, rotational excitation will show up merely as a broadening of the vibrational bands and the band shifts of vibrational transitions due to the coupling can be considered as being much smaller than the accuracy achievable by the quantum chemistry and quantum dynamics methods for obtaining the spectra. Therefore the following considerations will assume that rotational motion can be neglected, that is, we 
will set $\hat{\bfJ}=0$ or $\hat{P}_{\vartheta}=\hat{P}_{\phi}=\hat{P}_{\chi}=0$.
%
\section{\label{sec:app}Applications}
In the following we will give two applications of the present approach. First, we will consider the full-dimensional KEO for H$_3$O$_2^-$ employing polyspherical coordinates as introduced in more detail in Appendix \ref{keogatti}. Second, we consider N$_2$H$_7^+$ under the assumption of certain constraints and using bond length and angle coordinates.
\subsection{Full Dimensional KEO for H$_3$O$_2^-$}
%
In this section we will give the full dimensional KEO for the monohydrated hydroxide ion, H$_3$O$_2^-$, which had been used in the study of different isotopomers  in Ref. \cite{yang08:yyy}. The coordinates are defined in the MRF given by \Eq{LRF2MRF}. The four Jacobi vectors shown in \Fig{fig:h3o2-coords} are adopted. The three Euler angles are chosen in the same way as introduced in Section \ref{keogatti}, \ie, the MRF spherical coordinates $\bR_4$ and $\bR_1$ are $(R_4,0,0)$ and $(R_1,\theta_1,0)$, respectively. 
The other two vectors can be characterised by their spherical coordinates in the MRF $(R_j,\theta_j,\varphi_j)$ ($j=2,3$).
We can calculate the $\bG$ matrix according to \Eq{Minvs} and the required matrix elements are shown in \Eq{kevib}. Therefore,
we can directly write down the exact 9D KEO for total angular momentum $\bfJ=0$ as $2 \hat{T}_{\rm vib}=\hat{\bP}_{\rm vib}^\dagger \bG_{\rm vib}\hat{\bP}_{\rm vib}$ with the HCMOs given in Section \ref{hcmo}. However, this KEO will contain a large number of terms describing the angular momentum coupling between the shared Hydrogen and the (OH)$_{2}$ fragments. In order to simplify the KEO we will derive an approximate KEO which has the advantage that these couplings do not appear.

The basic idea is to introduce a new MRF and express $\bR_3$ with Cartesian coordinates $(x,y,z)$.
The remaining six coordinates are spherical coordinates in the old MRF $(R_1, R_2, \theta_1,\theta_2, R_4, \varphi=\varphi_2)$.
The new MRF is associated with the old MRF by a rotation of an angle $\eta\varphi$ around the $\bR_4$ (the $z$-axis).
The spherical coordinates for $\bR_3$ in the new MRF are $(R_3, \theta_3,\tilde{\varphi_3})$ with $
\tilde{\varphi_3}=\varphi_3-\eta\varphi$,
where $\eta=\mu_1/(\mu_1+\mu_2)$ is defined to minimise the Coriolis type couplings involving the central Hydrogen.
Here, $\mu_1$ ($\mu_2$) is the reduced mass associated with the Jacobi vector $\bR_1$ ($\bR_2$).
After transforming the spherical coordinates $(R_3, \theta_3,\tilde{\varphi_3})$ to the Cartesian ones $(x,y,z)$ we can obtain the KEO for the central Hydrogen
\be
T_3=-\frac{\hbar^2}{2\mu_3}\left(
\frac{\partial^2}{\partial x^2}+\frac{\partial^2}{\partial y^2}+\frac{\partial^2}{\partial z^2}
\right), 
\ee
where $\mu_3$ is the reduced mass associated with the Jacobi vector $\bR_3$ and Coriolis type terms are ignored.
Since the  $\bfe_z'$ is defined along the direction of $\bR_4$, the $z$ coordinate corresponds to the shared Hydrogen stretching vibration.

For the other 6 DOFs of the  (OH)$_2$ fragment, the 6D KEO is written in terms of the $\bG_{\rm vib}$ matrix elements \Eq{kevib}.
Here we take the angular momentum of the (OH)$_2$ fragment as total angular momentum since the angular momentum of the shared Hydrogen is negligible. Combining both parts we finally obtain the 9D KEO where we introduce
the new coordinates $u_i=\cos\theta_i$ ($i=1,2$).
For the  simulation an additional normalisation transform was performed to  reduce the numerical effort.
This gives the following 9D KEO:
\begin{equation}
T=T_{1}+T_{2}+T_{3}
\end{equation}
with
\begin{eqnarray}
T_{1}&=&-\frac{\hbar ^2}{2 \mu_1} \frac{\partial ^2}{\partial R_1^2} -\frac{\hbar ^2}{2 \mu_2} \frac{\partial ^2}{\partial R_2^2} -\frac{\hbar ^2}{2 \mu_4} \frac{\partial ^2}{\partial R_4^2}
\end{eqnarray}
\begin{eqnarray}
T_{2}&=&  -\sum_{i=1,2}\left(\frac{1}{2 \mu_i R_i^2}+\frac{1}{2 \mu_4 R_4^2}\right) \frac{\partial}{\partial u_i} (1-u_i^2) \frac{\partial}{\partial u_i} \nl
&&-\sum_{i=1,2}\left(\frac{1}{2 \mu_i R_i^2} \frac{1}{1-u_i^2}+ \frac{1}{2 \mu_4 R_4^2} \frac{u_i^2}{1-u_i^2}\right) \frac{\partial ^2}{\partial \varphi^2} \nl
&&+\frac{1}{ \mu_4 R_4^2} \frac{u_1}{\sqrt{1-u_1^2}} \frac{u_2}{\sqrt{1-u_2^2}} \frac{\partial }{\partial \varphi} \cos \varphi  \frac{\partial }{\partial \varphi} \nl
&&-\frac{1}{2 \mu_4 R_4^2}\left(\sqrt{1-u_1^2} \frac{\partial}{\partial u_1} \frac{\partial}{\partial u_2} \sqrt{1-u_2^2} + \frac{\partial}{\partial u_1} \sqrt{1-u_1^2} \sqrt{1-u_2^2} \frac{\partial}{\partial u_2} \right) \nl
&&-\frac{1}{2 \mu_4 R_4^2} \frac{u_2}{\sqrt{1-u_2^2}} \left(\frac{\partial }{\partial \varphi} \sin \varphi \sqrt{1-u_1^2} \frac{\partial }{\partial u_1}+ \frac{\partial }{\partial u_1} \sqrt{1-u_1^2}  \sin \varphi \frac{\partial }{\partial \varphi}\right)\nl
&&-\frac{1}{2 \mu_4 R_4^2} \frac{u_1}{\sqrt{1-u_1^2}} \left(\frac{\partial }{\partial \varphi} \sin \varphi \sqrt{1-u_2^2} \frac{\partial }{\partial u_2}+ \frac{\partial }{\partial u_2} \sqrt{1-u_2^2}  \sin \varphi \frac{\partial }{\partial \varphi}\right)\, .
\end{eqnarray}
The non-Euclidean normalisation according to the volume element is $d_\tau = dR_1 dR_2 dR_4 dx dy dz du_1 du_2 d \varphi$.
The reduced masses for different isotopomers are defined as follows: HOHOH$^-$ -- $\mu_1=\mu_2=m_{\rm H}m_{\rm O}/(m_{\rm H}+m_{\rm O})$, $\mu_3=2m_{\rm H}(m_{\rm H}+m_{\rm O})/(3m_{\rm H}+2m_{\rm O})$, $\mu_4=(m_{\rm H}+m_{\rm O})/2$; HODOH$^-$ -- $\mu_3=2m_{\rm D}(m_{\rm H}+m_{\rm O})/(m_{\rm D}+2m_{\rm H}+2m_{\rm O})$;  HOHOD$^-$ --
$\mu_1=m_{\rm  D} m_{\rm O}/(m_{\rm H}+m_{\rm O})$ and $\mu_3$ and $ \mu_4$ change correspondingly.
For DODOD$^-$ the corresponding  masses of HOHOH$^-$ are modified by replacing $m_{\rm H}$ by $m_{\rm D}$.
In the same way we get the masses for DOHOD$^-$ by exchanging $m_{\rm H}$ and $m_{\rm D}$ in HODOH$^-$, and similarly one can obtain
DODOH$^-$ from DOHOH$^-$.
%
\subsection{Reduced Dimensional KEO for N$_2$H$_7^+$}
In this section we give an example for a KEO for a system with constraints stressing again that  our method is invariant for reduced dimensional descriptions. Specifically we focus on  the protonated ammonia dimer, N$_2$H$_7^+$, whose IR spectrum in the range $<$ 1500 \cm had been discussed in Ref. \cite{asmis07:8691} on the basis of a 5D model including only the asymmetric stretching vibration of the shared proton. Subsequently, the model was extended to account for the two degenerate bendings as well in Ref. \cite{yang:}. Here we will derive the 5D model and comment on its extension.
A reasonable reduced description for the spectrum which is influenced by the shared proton motion should take into account the following two conditions: (i) experimental data suggest that the relevant energy range is presumably below 1000 \cm, but might extend into the $< 1500$ \cm range due to combination bands and  (ii) symmetry selection rules dominate the anharmonic couplings especially in this low-energy range.
Therefore, we first assume that the C$_3$ symmetry of the N$_2$H$_6$ fragment, i.e., excluding the central proton, will not be broken.  Second,  the length of the N-H covalent bonds shall be fixed. Notice that this neglects  the bending motion of the shared proton perpendicular to the line connecting the two nitrogen atoms. This is justified by the different symmetry and the lower anharmonicity of these two modes as compared to the proton motion along the N-N line.
These constraints leave five internal coordinates to describe the system as shown in Fig. \ref{fig:n2h7-coords}, i.e., the shared proton stretching with respect to the center of mass of the rest N$_2$H$_6$ fragment, $z$, the relative motion of the centres of mass of the two ammonia, $R$, the umbrella type motion of the two ammonia, $\theta_{1}$ and $\theta_{2}$, and the rotation (torsion) of the NH$_{3}$ fragments with respect to each other, $\varphi$.

For these five active coordinates the classical kinetic energy  with all other internal coordinates frozen can be obtained by exploiting \Eq{part-KE} repeatedly
 \bea
T&=&T_{(\mathrm{H})}^{(\mathrm{N}_2\mathrm{H}_6)}+T^{(\mathrm{N}_2\mathrm{H}_6)}_{\mathrm{N}_2\mathrm{H}_6} \nl
&=&T_{(\mathrm{H})}^{(\mathrm{N}_2\mathrm{H}_6)}+T^{(\mathrm{NH}_3)}_{\mathrm{NH}_3}+T^{(\mathrm{NH}_3')}_{\mathrm{NH}_3'}+T_{(\mathrm{NH}_3)}^{(\mathrm{NH}_3')} \nl
&=&T_{(\mathrm{H})}^{(\mathrm{N}_2\mathrm{H}_6)}+T_{(\mathrm{NH}_3)}^{(\mathrm{NH}_3')}+\left(T^{(\mathrm{H}_3)}_{\mathrm{H}_3}+T^{(\mathrm{H}_3)}_{(\mathrm{N})}\right)+\left(T^{(\mathrm{H}_3')}_{\mathrm{H}_3'}+T^{(\mathrm{H}_3')}_{(\mathrm{N}')}\right), 
\eea
where $(\mathrm{AB})$ is the center of mass of $\mathrm{AB}$. 
It is straightforward to obtain each term in above equation
  \bea
  \label{eq:tsep}
&&T_{(\mathrm{H})}^{(\mathrm{N}_2\mathrm{H}_6)}={\frac{1}{2}} \mu_p \dot{z} ^2 \nl
&&T_{(\mathrm{NH}_3)}^{(\mathrm{NH}_3)}={\frac{1}{2}}  \mu_R \dot{R} ^2 \nl
&&T^{(\mathrm{H}_3)}_{\mathrm{H}_3}={\frac{3}{2}} m_\mathrm{H} \left({\frac{d(R_0 \mbox{sin} \theta_{i}) }{dt}}\right) ^2+
{\frac{3}{2}} m_\mathrm{H} (R_0 \mbox{sin} \theta_{i}) ^2 \dot{\varphi}_{i} ^2 \nl
&&T^{(\mathrm{H}_3)}_{(\mathrm{N})}={\frac{1}{2}} \mu_{(\mathrm{N}-3\mathrm{H})} \left({\frac{d(R_0 \mbox{cos} \theta_{i}) }{dt}}\right) ^2, 
\eea
where $\mu_p=2m_\mathrm{H}(3m_\mathrm{H}+m_\mathrm{N})/(7m_\mathrm{H}+2m_\mathrm{N})$,
$\mu_R=\frac{1}{2}(3m_\mathrm{H}+m_\mathrm{N})$ and $\mu_{(\mathrm{N}-3\mathrm{H})}=3m_\mathrm{H}m_\mathrm{N}/(3m_\mathrm{H}+m_\mathrm{N})$.
The orientation angle of each individual ammonia is denoted by $\varphi_{1,2}$ and only the difference between them is the torsion shown in \Fig{fig:n2h7-coords} and $R_0$ is the free N-H covalent bond length. Thus the final kinetic energy is
  \bea
T&=&{\frac{1}{2}} \mu_p \dot{z} ^2
+ \mu _R \dot{R} ^2+
{\frac{3m_\mathrm{H} m_\mathrm{N}}{2(3m_\mathrm{H}+m_\mathrm{N})}} R_0^2 \left(\dot{\theta}_1^2 \mbox{sin} ^2 \theta_1+
\dot{\theta}_2^2 \mbox{sin} ^2 \theta_2\right) \nl
&&+{\frac{3}{2}} m_\mathrm{H} R_0^2 \sum_{i=1,2}(\dot{\theta}_i^2 \mbox{cos} ^2 \theta_i+\dot{\varphi}_i^2 \mbox{sin} ^2 \theta_i) \nl
&=&{\frac{1}{2}} \mu_p \dot{z} ^2
+{\frac{1}{2}}  \mu _R \dot{R} ^2+
{\frac{1}{2}} I_{\rm vib}({\theta}_1) \dot{\theta}_1 ^2+{\frac{1}{2}} I_{\rm vib}({\theta}_2) \dot{\theta}_2 ^2 \nl
&&+{\frac{1}{2}} I_{\rm rot}({\theta}_1) \dot{\varphi}_1 ^2+{\frac{1}{2}} I_{\rm rot}({\theta}_2) \dot{\varphi}_2 ^2, 
\eea
where $I_{\rm vib}(\theta)=I_0 (\mbox{cos} ^2 \theta + m_\mathrm{N}\mbox{sin} ^2 \theta)/(3m_\mathrm{H}+m_\mathrm{N})$, $I_{\rm rot}({\theta})=I_0 \mbox{sin} ^2 \theta$ and $I_0=3m_\mathrm{H} R_0 ^2$.
The last two terms in above equation can be rewritten and after separation of the global rotation we finally simplify the kinetic energy as
\bea
T&=&\frac{1}{2} \mu_p \dot{z}^2
+\frac{1}{2} \mu_R \dot{R}^2
+\sum_{i=1,2}\frac{1}{2} I_{\rm vib}(\theta_i) \dot{\theta_i}^2
+\frac{1}{2} I_{\rm tor}(\theta_1,\theta_2) \dot{\varphi}^2,
\eea
where the reduced moment of inertia for the torsion is $I_{\rm tor}({\theta}_1,\theta_2)=I_{\rm rot}({\theta}_1) I_{\rm rot}({\theta}_2)/(I_{\rm rot}({\theta}_1)+I_{\rm rot}({\theta}_2))$.

So far we did not consider the Coriolis type couplings with total angular velocity. However, due to the $C_3$ symmetry, the total angular momentum equals to zero is equivalent to total angular velocity equals to zero for this specific case since none of the adopted coordinates contributes to total angular momentum. Therefore the Coriolis type couplings are zero.

According to the general procedure detailed in Section \ref{fulldim} we can obtain the following quantum KEO ($\varphi=\varphi_{2}-\varphi_{1}$)
\bea
\hat{T}&=&\frac{1}{2 \mu_p} \hat{P}_{z}^\dagger \hat{P}_{z}
+\frac{1}{2 \mu_R} \hat{P}_{R}^\dagger \hat{P}_{R} 
+\frac{\hat{P}_{\varphi}^\dagger \hat{P}_{\varphi}}{2 I_{\rm tor}(\theta_1,\theta_2)} \nl
&&+\frac{1}{2} \hat{P}_{\theta_1}^\dagger {I_{\rm vib}^{-1}(\theta_1)} \hat{P}_{\theta_1}
+\frac{1}{2} \hat{P}_{\theta_2}^\dagger {I_{\rm vib}^{-1}(\theta_2)} \hat{P}_{\theta_2}.
\eea
Inspecting the adopted coordinates we notice that one of them is Cartesian and the other four are spherical coordinates defined in MRF.
Therefore, according to Section \ref{hcmo} the HCMOs are given by
\bea
\hat{P}_{z}^\dagger&=&\hat{P}_{z} \nl
\hat{P}_{R}^\dagger&=&\hat{P}_{R}-\frac{2i\hbar}{R} \nl
\hat{P}_{\theta_j}^\dagger&=&\hat{P}_{\theta_j}-i\hbar \mbox{cot} \theta_j,j=1,2 \nl
\hat{P}_{\varphi}^\dagger&=&\hat{P}_{\varphi}. 
\eea
Hence the final KEO and  the  Euclidean normalisation condition are given as follows
\bea\label{eq:tkin}
\hat{T}&=&-\frac{\hbar ^2}{2 \mu_p} \frac{\partial ^2}{\partial z^2}
-\frac{\hbar ^2}{2 \mu_{_R}} \frac{1}{R^2} \frac{\partial}{\partial R} R^2 \frac{\partial}{\partial R} \nl
&&-\frac{\hbar ^2}{2 I_{\rm tor}(\theta_1,\theta_2)} \frac{\partial ^2}{\partial \varphi ^2}
-\frac{\hbar ^2}{2} \sum_{i=1,2} \frac{1}{\sin \theta_i} \frac{\partial}{\partial \theta_i} \frac{\sin \theta_i}{I_{\rm vib}(\theta_i)} \frac{\partial}{\partial \theta_i} \nl
d \tau&=&R^2 \sin \theta_1 \sin \theta_2 dz dR d \theta_1 d \theta_2 d \varphi. 
\eea

In Ref. \cite{yang:} we have extended this model by including the shared proton bending coordinates $x$ and $y$ as shown in Fig. \ref{fig:n2h7-coords}. In terms of the KEO this is straightforward since in \Eq{eq:tsep} we simply have to set
\begin{equation}
T_{(\mathrm{H})}^{(\mathrm{N}_2\mathrm{H}_6)}={\frac{1}{2}} \mu_p (\dot{z} ^2 
+ \dot{x} ^2+ \dot{y} ^2 )
\end{equation}
which is carried through the derivation to give a corresponding contribution to \Eq{eq:tkin}. Compared to the experiment this gives a rather good description of the bending fundamental transitions as well as an improvement concerning the shared proton fundamental stretching transition. 
\section{\label{sec:concl}Summary}
In summary we have presented a new method for obtaining  kinetic energy operators which involves a straightforward quantisation of the classical Lagrange/Hamilton function. It is based on the notion of hermitian conjugate momentum operators whose derivation presents the major effort for practical applications. An important point of our approach is its validity for systems with constraints. This makes it particularly attractive for studying larger systems in reduced dimensionality. 

We have applied our approach to the case of a general system described by polyspherical coordinates and showed the equivalence of the kinetic energy operator with that obtained by the method of Gatti and coworkers. Afterwards we discussed two specific applications to charged cluster systems having strong hydrogen bonds. The strong coupling between different coordinates as well their structural floppiness makes it necessary to perform high dimensional quantum dynamics simulations as done in Refs. \cite{asmis07:8691,yang08:yyy,yang:} on the basis of the kinetic energy operators derived in Section \ref{sec:app}.

\section{Appendix}
\subsection{Hermitian Conjugate Momentum Operators for Spherical Coordinates}
\label{sec:appd}
The derivation of the expressions for the HCMOs in LRF is the most tedious part of the present method and it becomes even more involved if 
coordinates defined in MRFs are used.  In this Appendix we will first summarise the relevant relations between the  LRF and MRFs and subsequently turn to the derivation of HCMOs. 

Consider the LRF defined by three orthogonal unit vectors $\{\bfe_x, \bfe_y, \bfe_z\}$ and a MRF defined by three orthogonal unit vectors $\{\bfe_{x'}, \bfe_{y'}, \bfe_{z'}\}$.
The orientation angles of $\bfe_{z'}$ in the LRF are $(\vartheta,\phi)$.
To obtain the connection between the LRF and the MRF we first apply two excessive rotations $\bU_y(\vartheta)$ and $\bU_z(\phi)$ to the LRF, where $\bU_y(\theta)$ means rotating $\vartheta$ around $\bfe_{y}$ and $\bU_z(\phi)$ means rotating $\phi$ around $\bfe_{z}$.
The matrix representation of them are
\bea
\bU_y(\vartheta)&=&
\left(
\begin{array}{ccc}
\cos\vartheta & 0 & \sin\vartheta\\
0 & 1 & 0\\
-\sin\vartheta & 0 & \cos\vartheta
\end{array}
\right) \nl
\bU_z(\phi)&=&
\left(
\begin{array}{ccc}
\cos\phi & -\sin\phi & 0\\
\sin\phi & \cos\phi & 0\\
0 & 0 & 1
\end{array}
\right).
\eea
The new reference frame generated by applying $\bU_y(\vartheta)$ and $\bU_z(\phi)$ to the LRF has the same $z$ axis as the MRF, \ie, the only difference between the two reference frames is just a rotation by $\chi$ around $\bfe_{z'}$.
Therefore, the MRF can be obtained by applying three successive rotations $\bU_y(\vartheta)$, $\bU_z(\phi)$ and $\bU_{z'}(\chi)$ to the LRF, namely
\be\label{aatrans1}
\bfe_{\alpha'} = \bU_{z'}(\chi) \bU_z(\phi) \bU_y(\vartheta) \bfe_{\alpha}, \hspace{5mm}\alpha=x,y,z,
\ee 
where $\bU_{z'}(\chi)$ means rotating $\chi$ around $\bfe_{z'}$.
The matrix representation for $\bU_{z'}(\chi)$ in the MRF is the same with $\bU_{z}(\chi)$ in the LRF.
Since the third rotation $\bU_{z'}(\chi)$ in \Eq{aatrans1} has no effects on $\bfe_{z'}$ we actually have $\bfe_{z'} =\bU_z(\phi) \bU_y(\vartheta) \bfe_{z}$.
Exploiting the rules between vector and operator transformations we can obtain the expression for $\bU_{z'}(\chi)$ in the LRF
\be
\bU_{z'}(\chi)=\left[\bU_z(\phi) \bU_y(\vartheta) \right] \bU_{z}(\chi)
\left[\bU_z(\phi) \bU_y(\vartheta) \right]^{-1}.
\ee
Consequently \Eq{aatrans1} can be rewritten as
\be\label{aatrans}
\bfe_{\alpha'} = \bU_z(\phi) \bU_y(\vartheta)\bU_{z}(\chi)  \bfe_{\alpha} \hspace{5mm}\alpha=x,y,z.
\ee 
\Eq{aatrans} tells us that an equivalent way to obtain the MRF is to apply three successive rotations $\bU_{z}(\chi)$, $\bU_y(\vartheta)$ and $\bU_z(\phi)$ to the LRF.

Now let us consider vectors $\bR_j$ characterised by spherical coordinates $(R_j, \theta_j, \varphi_j)$ in the MRF ($j=1-N$).
With \Eq{aatrans} and the rule between basis vectors and components transformations we can obtain the Cartesian components of these vectors in LRF
\be\label{coortrans}
\left(
\begin{array}{c}
R_{jx} \\
R_{jy}\\
R_{jz}
\end{array}
\right)
=\bU_z(\phi) \bU_y(\vartheta) \bU_z(\chi)
\left(
\begin{array}{c}
R_{jx'} \\
R_{jy'}\\
R_{jz'}
\end{array}
\right) ,
\ee
where $R_{j\alpha}$ and $R_{j\alpha'}$ are Cartesian components of $\bR_j$ in the LRF and the MRF, respectively.
We can rewrite \Eq{coortrans} in more formally as
\be\label{aacoortrans1}
\bR_j=R_j \bU_z(\phi) \bU_y(\vartheta) \bU_z(\chi) \bU_z(\varphi_j) \bU_y(\theta_j) \bfe_z,
\ee
\Eq{aacoortrans1} is a vector equation therefore it is also valid for an arbitrary reference frame.

Now we turn to the main task, that is, deriving the Hermitian conjugate of momentum operators. 
This can be done step by step according to \Eq{pdgmore} with the help of the coordinate transformation \Eq{coortrans}. 
From \Eq{coortrans} we can see that $(\theta_j, \varphi_j)$ only appear in the Cartesian components of $\bR_j$. 
This greatly simplifies the expression for the Hermitian conjugates of momentum operators as follows
\bea\label{momentumhc}
\hat{P}_{R_j}^{\dagger}&=&\hat{P}_{R_j}-i \hbar
\sum_{\alpha=x,y,z} \left(\frac{\partial}{\partial R_{j \alpha}} \frac{\partial
R_{j \alpha}}{\partial R_j}\right)^\circ \nl
\hat{P}_{\theta_j}^{\dagger}&=&\hat{P}_{\theta_j}-i \hbar
\sum_{\alpha=x,y,z} \left(\frac{\partial}{\partial R_{j \alpha}} \frac{\partial
R_{j \alpha}}{\partial \theta_j}\right)^\circ \nl
\hat{P}_{\varphi_j}^{\dagger}&=&\hat{P}_{\varphi_j}-i \hbar
\sum_{\alpha=x,y,z} \left(\frac{\partial}{\partial R_{j \alpha}} \frac{\partial
R_{j \alpha}}{\partial \varphi_j}\right)^\circ.
\eea

Here  the three Euler angles in \Eq{coortrans} are purely parameters.
To calculate the partial derivatives we first give some useful relations following from the orthogonality of the transformation \Eq{coortrans}
\bea\label{ctrans3}
&&R_{j \alpha}=\sum_{\beta'} \frac{\partial R_{j \alpha}}{\partial R_{j \beta'}}R_{j \beta'}, \hspace{10mm}
R_{j \beta'}=\sum_{\alpha} \frac{\partial R_{j \beta'}}{\partial R_{j \alpha}}R_{j \alpha} \nl
&&\sum_{\beta'} \left(\frac{\partial R_{j\alpha}}{\partial R_{j \beta'}}\right)^2=
\sum_{\alpha} \left(\frac{\partial R_{j\alpha}}{\partial R_{j \alpha'}}\right)^2=1
\nl
&&\frac{\partial R_{j\alpha}}{\partial R_{j \beta'}}=\frac{\partial R_{j\beta'}}{\partial R_{j \alpha}}
\eea
where $\alpha=x,y,z$ and $\beta'=x',y',z'$.
Note here all the derivatives $\{\partial R_{j \beta'}/\partial R_{j \alpha}\}$ are parameters only depending on the three Euler angles.

Let us first derive the expression for $\hat{P}_{R_j}^{\dagger}$. It is straightforward to get the following derivatives of MRF Cartesian coordinates
\bea\label{deriv1}
\frac{\partial R_{jx'}}{\partial R_j}&=& \sin\theta_j\cos\varphi_j
=\frac{R_{jx'}}{R_{j}}\nl
\frac{\partial R_{jy'}}{\partial R_j}&=& \sin\theta_j\sin\varphi_j
=\frac{R_{jy'}}{R_{j}}\nl
\frac{\partial R_{jz'}}{\partial R_j}&=& \cos\theta_j
=\frac{R_{jz'}}{R_{j}}\,
\eea
where $R_j^2=R_{jx}^2+R_{jy}^2+R_{jz}^2=R_{jx'}^2+R_{jy'}^2+R_{jz'}^2$.
Based on \Eq{deriv1} we can obtain the derivatives of LRF Cartesian coordinates with respect to $R_j$ and derive the final result
\bea
\sum_{\alpha} \left(\frac{\partial}{\partial R_{j \alpha}} \frac{\partial
R_{j \alpha}}{\partial R_j}\right)^\circ 
&=&\sum_{\alpha} \left(\frac{\partial}{\partial R_{j \alpha}} \sum_{\alpha'}
\frac{R_{j\beta'}}{R_j} \frac{\partial R_{j \alpha}}{\partial R_{j\beta'}}\right)^\circ \nl
&=&
\sum_{\alpha,\beta'}\left(\frac{1}{R_j}\frac{\partial R_{j\beta'}}{\partial R_{j \alpha}}+
R_{j\beta'}\frac{\partial \left(R_{jx}^2+R_{jy}^2+R_{jz}^2\right)^{-\frac{1}{2}} }{\partial R_{j \alpha}}\right)\frac{\partial R_{j \alpha}}{\partial R_{j\beta'}}\nl
&=&
\sum_{\alpha,\beta'}
\left(\frac{1}{R_j}\frac{\partial R_{j\beta'}}{\partial R_{j \alpha}}
-\frac{R_{j\beta'}R_{j\alpha}}{R_j^3}\right)\frac{\partial R_{j\alpha}}{\partial R_{j \beta'}} \nl
&=&\sum_{\alpha}\left(\frac{1}{R_j}
-\frac{R_{j\alpha}^2}{R_j^3}\right) \nl
&=&\frac{3}{R_j}-\frac{1}{R_j}=\frac{2}{R_j},
\eea
where \Eq{ctrans3} has been used in the fourth step.
Finally we can get the HCMO associated with $R_j$ as follows
\be\label{rdg12}
\hat{P}_{R_j}^{\dagger}=\hat{P}_{R_j}-\frac{2i\hbar}{R_j}
=-i\hbar\frac{1}{R_j^2}\frac{\partial}{\partial R_j}R_j^2.
\ee

Next we will derive the expression for $\hat{P}_{\theta_j}^{\dagger}$ following the same procedure. The derivatives of MRF Cartesian coordinates read
\bea\label{thderiv1}
\frac{\partial R_{jx'}}{\partial \theta_j}&=&R_j \cos\theta_j\cos\varphi_j
=R_{jx'}\cot\theta_j\nl
\frac{\partial R_{jy'}}{\partial \theta_j}&=&R_j \cos\theta_j\sin\varphi_j
=R_{jy'}\cot\theta_j\nl
\frac{\partial R_{jz'}}{\partial \theta_j}&=&-R_j\sin\theta_j=-\frac{R_{jz'}}{\cot\theta_j}.
\eea
Based on \Eq{thderiv1} we can obtain the derivatives of Cartesian coordinates in LRF with respect to $\theta_j$
\be\label{thderiv2}
\frac{\partial R_{j \alpha}}{\partial \theta_j}=
R_{jx'}\cot\theta_j\frac{\partial R_{j \alpha}}{\partial R_{jx'}}
+R_{jy'}\cot\theta_j\frac{\partial R_{j \alpha}}{\partial R_{jy'}}
-\frac{R_{jz'}}{\cot\theta_j}\frac{\partial R_{j \alpha}}{\partial R_{jz'}}
\ee
Furthermore, we obtain
\bea\label{thderiv3}
\left(\frac{\partial }{\partial R_{j \alpha}} \frac{\partial R_{j \alpha}}{\partial \theta_j}\right)^\circ
&=&\cot\theta_j\frac{\partial R_{j \alpha}}{\partial R_{jx'}}
\frac{\partial R_{jx'}}{\partial R_{j \alpha}}
+\cot\theta_j\frac{\partial R_{j \alpha}}{\partial R_{jy'}}
\frac{\partial R_{jy'}}{\partial R_{j \alpha}}
-\frac{1}{\cot\theta_j}\frac{\partial R_{j \alpha}}{\partial R_{jz'}}
\frac{\partial R_{jz'}}{\partial R_{j \alpha}} \nl
&&+\left(
R_{jx'}\frac{\partial R_{j \alpha}}{\partial R_{jx'}}
+R_{jy'}\frac{\partial R_{j \alpha}}{\partial R_{jy'}}
+\frac{R_{jz'}}{\cot^2\theta_j}\frac{\partial R_{j \alpha}}{\partial R_{jz'}}
\right)\frac{\partial \cot\theta_j}{\partial R_{j \alpha}} \nl
&=&\cot\theta_j-\frac{1}{\sin\theta_j\cos\theta_j}
\frac{\partial R_{j \alpha}}{\partial R_{jz'}}
\frac{\partial R_{jz'}}{\partial R_{j \alpha}} \nl
&&+\left(
R_{j\alpha}
+(\tan^2\theta_j-1)R_{jz'}\frac{\partial R_{j \alpha}}{\partial R_{jz'}}
\right)\frac{\partial \cot\theta_j}{\partial R_{j \alpha}}.
\eea
Since $\cot\theta_j=R_{jz'}/\sqrt{R_{jx'}^2+R_{jy'}^2}$ we can get the following derivatives
\bea\label{aathderv}
\frac{\partial \cot\theta_j}{\partial R_{jx'}}&=&-\frac{R_{jx'}R_{jz'}}{\sqrt{R_{jx'}^2+R_{jy'}^2}^3} \nl
\frac{\partial \cot\theta_j}{\partial R_{jy'}}&=&-\frac{R_{jy'}R_{jz'}}{\sqrt{R_{jx'}^2+R_{jy'}^2}^3} \nl
\frac{\partial \cot\theta_j}{\partial R_{jz'}}&=&\frac{1}{\sqrt{R_{jx'}^2+R_{jy'}^2}}.
\eea
Based on \Eq{aathderv} we have
\bea\label{aathdv1}
\frac{\partial \cot\theta_j}{\partial R_{j \alpha}}
&=&
\frac{1}{\sqrt{R_{jx'}^2+R_{jy'}^2}^3}\left(
-R_{jx'}R_{jz'}\frac{\partial R_{jx'}}{\partial R_{j \alpha}}
-R_{jy'}R_{jz'}\frac{\partial R_{jy'}}{\partial R_{j \alpha}}
+(R_{jx'}^2+R_{jy'}^2)\frac{\partial R_{jz'}}{\partial R_{j \alpha}}
\right) \nl
&=&\frac{1}{\sqrt{R_{jx'}^2+R_{jy'}^2}^3}\left(
-R_{jz'}R_{j \alpha}+R_j^2\frac{\partial R_{jz'}}{\partial R_{j \alpha}}
\right).
\eea
Combining \Eq{thderiv3} and \Eq{aathdv1} we can now derive
\bea\label{finalhc}
&&\sum_{\alpha}\left(\frac{\partial }{\partial R_{j \alpha}}\frac{\partial R_{j \alpha}}{\partial \theta_j}\right)^\circ \nl
&=&\sum_{\alpha}\left(
\cot\theta_j-\frac{1}{\sin\theta_j\cos\theta_j}
\frac{\partial R_{j \alpha}}{\partial R_{jz'}}
\frac{\partial R_{jz'}}{\partial R_{j \alpha}}\right) \nl
&+&\sum_{\alpha}\frac{1}{\sqrt{R_{jx'}^2+R_{jy'}^2}^3}\left(
R_{j\alpha}
+(\tan^2\theta_j-1)R_{jz'}\frac{\partial R_{j \alpha}}{\partial R_{jz'}}
\right)\left(
-R_{jz'}R_{j \alpha}+R_j^2\frac{\partial R_{jz'}}{\partial R_{j \alpha}}
\right) \nl
&=&3\cot\theta_j-\frac{1}{\sin\theta_j\cos\theta_j} \nl
&+&\frac{1}{\sqrt{R_{jx'}^2+R_{jy'}^2}^3}\left(
-R_{jz'}R_{j}^2+R_{j}^2R_{jz'}+(\tan^2\theta_j-1)(-R_{jz'}^2R_{jz'}+R_{j}^2R_{jz'})\right)\nl
&=&3\cot\theta_j-\frac{1}{\sin\theta_j\cos\theta_j}
+(\tan^2\theta_j-1)\cot\theta_j \nl
&=&\cot\theta_j
\eea
The Hermitian conjugate of each $\hat{P}_{\theta_j}$ associated with the corresponding MRF polar angle $\theta_j$ can be finally expressed as
\be\label{finalpthc}
\hat{P}_{\theta_j}^{\dagger}=\hat{P}_{\theta_j}-i \hbar \cot \theta_j
=-\frac{i\hbar}{\sin\theta_j} \frac{\partial}{\partial \theta_j} \sin\theta_j.
\ee

Finally we come to $\hat{P}_{\varphi_j}^{\dagger}$. 
The procedure is the same, however, it is much simpler as  compared to $\hat{P}_{\theta_j}^{\dagger}$. 
Again we  start with the derivatives of MRF Cartesian coordinates
\bea
\frac{\partial R_{jx'}}{\partial \varphi_j}&=&-R_j \sin\theta_j\sin\varphi_j
=-R_{jy'} \nl
\frac{\partial R_{jy'}}{\partial \varphi_j}&=&R_j \sin\theta_j\cos\varphi_j
=R_{jx'}\nl
\frac{\partial R_{jz'}}{\partial \varphi_j}&=&0
\eea 
to obtain the final
\bea\label{pphihc}
\left(\sum_{\alpha}\frac{\partial }{\partial R_{j \alpha}}\frac{\partial R_{j \alpha}}{\partial \varphi_j}\right)^\circ 
&=&
\left(\sum_{\alpha}\frac{\partial }{\partial R_{j \alpha}}
\left(R_{jx'}\frac{\partial R_{j \alpha}}{\partial R_{jy'}}
-R_{jy'}\frac{\partial R_{j \alpha}}{\partial R_{jx'}}\right)\right)^\circ\nl
&=&
\sum_{\alpha}\left(
\frac{\partial R_{jx'}}{\partial R_{j \alpha}}\frac{\partial R_{j\alpha}}{\partial R_{j y'}}
-\frac{\partial R_{jy'}}{\partial R_{j \alpha}}\frac{\partial R_{j\alpha}}{\partial R_{j x'}}
\right)=0
\eea
The momentum operator associated with each orientation angle $\varphi_j$ is Hermitian according to \Eq{pphihc}, namely
\be\label{phidg11}
\hat{P}_{\varphi_j}^{\dagger}=\hat{P}_{\varphi_j}=-i\hbar \frac{\partial}{\partial\varphi_j}.
\ee

Let us recall the above detailed procedure. The only condition we need is that the transformation matrix between Cartesian coordinates in the LRF and those in a MRF is orthogonal. First, if we set all the three Euler angles equal to zero we can obtain the HCMOs associated with spherical coordinates in the LRF. The final results are the same with \Eq{rdg12}, \Eq{finalpthc}, and \Eq{phidg11} since a unit matrix is also an orthogonal matrix.
Second, we can use more rotations to define more MRFs and the HCOMs associated with the spherical coordinates in each different MRF obey \Eq{rdg12}, \Eq{finalpthc}, and \Eq{phidg11}.
As an important consequence the Hermitian conjugates of all the momentum operators associated with \emph{real} bond lengths, bond angles and dihedral angles obey \Eq{rdg12}, \Eq{finalpthc}, and \Eq{phidg11}, respectively, irrespective of how complicated the molecule might be.
Similarly, the result holds for any reference frame provided there exists an orthogonal transformation to transform it to the LRF.
Based on the above conclusions we can see the HCMOs associated with the three Euler angles obey the same relations
\bea
\hat{P}_{\vartheta}^\dagger&=&\hat{P}_{\vartheta}-i\hbar\cot\vartheta
=-\frac{i\hbar}{\sin\vartheta} \frac{\partial}{\partial \vartheta} \sin\vartheta \nl
\hat{P}_{\phi}^{\dagger}&=&\hat{P}_{\phi}=-i\hbar \frac{\partial}{\partial\phi} \nl
\hat{P}_{\chi}^{\dagger}&=&\hat{P}_{\chi}=-i\hbar \frac{\partial}{\partial\chi}.
\eea

Finally, we can obtain the HCMOs $\hat{P}_{j\alpha'}^\dagger$ associated with MRF Cartesian coordinates following the same procedure.
Since $\partial R_{j \alpha}/\partial R_{j \alpha'}$ is just a parameter which does not depends on $R_{jx}$, $R_{jy}$ or $R_{jz}$ we can easily get
\be
\label{eq:cartHerm}
\hat{P}_{j\alpha'}^\dagger=\hat{P}_{j\alpha'}-i \hbar
\sum_{\alpha} \left(\frac{\partial}{\partial R_{j \alpha}} \frac{\partial
R_{j \alpha}}{\partial R_{j \alpha'}}\right)^\circ =\hat{P}_{j\alpha'},
\ee
where $\alpha'=x',y',z'$. Apparently, the momenta associated with MRF Cartesian coordinates are Hermitian.
%
\subsection{KEO in Terms of Polyspherical Coordinates}\label{keogatti}
%
In order to connect our approach to existing one we consider the KEO in terms of so called polyspherical coordinates defined in the LRF and a MRF. This will lead us to expression which have been reported in Ref. \cite{gatti01:8275}.
Consider a molecular system composed of $N+1$ atoms.
After separating the total centre of mass motion we can describe it with $N$ vectors $\bR_1$, $\bR_2$, $\cdots$, $\bR_N$.
The three Euler angles $(\vartheta, \phi, \chi)$ are chosen in such a way that the spherical coordinates of $\bR_N$ in the LRF are $(R_N,\vartheta,\phi)$ and the spherical coordinates of $\bR_1$ in the MRF are $(R_1,\theta_1,\varphi_1=0)$.
That is to say the $\bfe_{z'}$ axis of the MRF is defined to be along the direction of $\bR_N$.
The rest $N-2$ vectors are characterised by spherical coordinates $\{R_j,\theta_j,\varphi_j\}$ in the MRF $(j=2,\cdots,N-1)$.

Using \Eq{eqmf1} 
with $j=1,\cdots,N$ and $\theta_N=\varphi_N=\varphi_1=0$, we can derive the LRF components of the $N$ velocity vectors
\bea\label{acvelocity}
\dot{\bR}_j&=&
R_j\dot{\vartheta} \bU_z(\phi) \bU'_y(\vartheta) \bU_z(\chi) \bU_z(\varphi_j) \bU_y(\theta_j) \bfe_z \nl
&+& R_j\dot{\phi} \bU'_z(\phi) \bU_y(\vartheta)\bU_z(\chi) \bU_z(\varphi_j) \bU_y(\theta_j) \bfe_z \nl
&+& R_j\dot{\chi} \bU_z(\phi) \bU_y(\vartheta)\bU'_z(\chi) \bU_z(\varphi_j) \bU_y(\theta_j) \bfe_z \nl
&+&\dot{R}_j \bU_z(\phi) \bU_y(\vartheta) \bU_z(\chi) \bU_z(\varphi_j) \bU_y(\theta_j)\bfe_z\nl
&+&R_j\dot{\theta}_j \bU_z(\phi) \bU_y(\vartheta) \bU_z(\chi) \bU_z(\varphi_j) \bU'_y(\theta_j) \bfe_z\nl
&+&R_j\dot{\varphi}_j\bU_z(\phi) \bU_y(\vartheta) \bU_z(\chi) \bU'_z(\varphi_j) \bU_y(\theta_j) \bfe_z,
\eea
where $\bU'$ is the first order derivative of the corresponding rotational transformation matrix. The above equations are still vector equations, however, they are only valid in the LRF since we have used the relation $\dot{\bfe_z}=0$. 
To make them valid in an arbitrary reference frame we only need to add the corresponding terms containing $\dot{\bfe_z}$ to the right hand side.

Before proceeding we will discuss the orthogonality of the terms appearing in \Eq{acvelocity} .
The last three terms in \Eq{acvelocity} are just the spherical velocity components measured in the MRF while the first three terms are components of the velocity caused by the non-inertial MRF.
Each term in \Eq{acvelocity} can be expressed in a formally simple way in terms of angular velocity vectors
\bea\label{acv2}
\dot{\bR}_j&=&\dot{\vartheta}\bfe_{\dot{\vartheta}}\times\bR_j+
\dot{\phi}\bfe_{\dot{\phi}}\times\bR_j+
\dot{\chi}\bfe_{\dot{\chi}}\times\bR_j\nl
&+&\dot{R}_j\bR_j/R_j+
\dot{\theta_j}\bfe_{\dot{\theta_j}}\times\bR_j+
\dot{\varphi_j}\bfe_{\dot{\varphi_j}}\times\bR_j,
\eea
where $\bfe_{\dot{\vartheta}}$ is the direction of the angular velocity $\dot{\vartheta}$, and similar for $\bfe_{\dot{\phi}}$, $\bfe_{\dot{\chi}}$, $\bfe_{\dot{\theta}_j}$, and $\bfe_{\dot{\varphi}_j}$.
With the help of \Eq{acv2} one can immediately see that the fourth term is parallel to $\bR_j$ while the other terms are perpendicular to $\bR_j$.
Recalling the velocity in spherical coordinates we know that the last three terms are orthogonal to each other.
Therefore, in \Eq{acvelocity} (or \Eq{acv2} ), the fourth term (namely the $\dot{R}_j$ term) is orthogonal to the other terms and the last three terms are orthogonal to each other.

Having at hand all the velocities it is quite straightforward to write the classical kinetic energy \cite{iung99:3377} according to 
\be
T=\frac{1}{2}\sum_{i,j=1}^N \mu_{ij} \dot{\bR}_i^{\dagger} \dot{\bR}_j,
\ee 
where the matrix $\{\mu_{jk}\}$ combines the reduced masses and the transformation matrix between adopted  and Jacobi vectors \cite{iung99:3377}.
The final results in LRF can be written in a symmetric form as in \Eq{lagrange}
\be
T=\frac{1}{2} \dot{\bQ}^{\dagger} \bM \dot{\bQ}, \hspace{10mm}
\bQ=\left(
\begin{array}{c}
\bQ_{\rm rot} \\
\bQ_{\rm vib}
\end{array}
\right),
\ee
where the coordinates are separated as rotational, \Eq{eq:qrot}, and vibrational DOFs
\bea
\bQ_{\rm vib}^{\dagger}&=&\left(
\begin{array}{ccccccccc}
R_1 & \cdots & R_N & \theta_1 & \cdots & \theta_{N-1} & \varphi_2 & \cdots & \varphi_{N-1}
\end{array}
\right). 
\eea 

The next step is to calculate the inverse matrix of $\bM$ to generate the quantum KEO. 
Aiming at a separation of the rotational and vibrational DOFs, we divide the matrix $\bM$ into the following four blocks 
\be
\bM=\left(
\begin{array}{cc}
\bM_{\rm rot} & \bM_{\rm Cor}\\
\bM_{\rm Cor}^\dagger & \bM_{\rm vib}
\end{array}
\right),
\ee
where the subscripts ``$_{\rm rot}$'', ``$_{\rm Cor}$'', and ``$_{\rm vib}$'' correspond to rotational, Coriolis, and vibrational terms, respectively.
Therefore, $\bM_{\rm rot}$, $\bM_{\rm Cor}$, and $\bM_{\rm vib}$ are $3\times 3$, $3\times (3N-3)$ and $(3N-3)\times (3N-3)$ dimensional matrices, respectively.
Suppose the inverse matrix of $\bM$ is divided in the same spirit as
\be
\bM^{-1}=\left(
\begin{array}{cc}
\bG_{\rm rot} & \bG_{\rm Cor}\\
\bG_{\rm Cor}^\dagger & \bG_{\rm vib}
\end{array}
\right),
\ee
we can express the quantum KEO as a sum of rotational, vibrational, and Coriolis terms
\bea
\hat{T}&=&\hat{T}_{\rm rot}+\hat{T}_{\rm vib}+\hat{T}_{\rm Cor} \nl
&=&\frac{1}{2}\hat{\bP}_{\rm rot}^\dagger \bG_{\rm rot}\hat{\bP}_{\rm rot}+
\frac{1}{2}\hat{\bP}_{\rm vib}^\dagger \bG_{\rm vib}\hat{\bP}_{\rm vib}+
\frac{1}{2}\left(\hat{\bP}_{\rm rot}^\dagger \bG_{\rm Cor}\hat{\bP}_{\rm vib}+{\rm h.c.}\right).
\eea
According to Section \ref{angular}, the generalised momentum vector $\hat{\bP}_{\rm rot}$ associated with the rotational DOFs is just the total angular momentum vector $\bfJ$. If we are interested in the rotational DOFs only, we obtain upon setting the vibrational momenta equal to zero:
\be
\hat{T}_{\rm rot}=\frac{1}{2}\hat{\bP}_{\rm rot}^\dagger \bG_{\rm rot}\hat{\bP}_{\rm rot}
=\frac{1}{2}\hat{\bfJ}^\dagger \bG_{\rm rot}\hat{\bfJ},
\ee
where the components of $\bfJ$ should be along the directions of $\bfe_{\dot {\vartheta}}$, $\bfe_{\dot {\phi}}$ and $\bfe_{\dot {\chi}}$ according to \Eq{J-P1}. In passing we note that one can also transform the components to Cartesian ones according to \Eq{J-P}.
On the other hand, if we are interested in the vibrational DOFs, the KEO for the total angular momentum $\bfJ=0$ reads
\be
\hat{T}_{\rm vib}=\frac{1}{2}\hat{\bP}_{\rm vib}^\dagger \bG_{\rm vib}\hat{\bP}_{\rm vib}.
\ee

Finally, we  give the details on how to calculate the different blocks of the inverse matrix of $\bM$.
One can see that the matrix $\bM$ can be congruently block-diagonalised in the following way
\bea\label{acmatinrs}
&&\left(
\begin{array}{cc}
\textbf{1} & -\bM_{\rm Cor}\bM_{\rm vib}^{-1}\\
	\textbf{0} & \textbf{1}
\end{array}
\right)
\left(
\begin{array}{cc}
\bM_{\rm rot} & \bM_{\rm Cor}\\
\bM_{\rm Cor}^\dagger & \bM_{\rm vib}
\end{array}
\right)
\left(
\begin{array}{cc}
\textbf{1} & \textbf{0}\\
-\bM_{\rm vib}^{-1}\bM_{\rm Cor}^\dagger & \textbf{1}
\end{array}
\right) \nl
&&=
\left(
\begin{array}{cc}
\bM_{\rm rot}-\bM_{\rm Cor}\bM_{\rm vib}^{-1}\bM_{\rm Cor}^\dagger & \textbf{0}\\
\textbf{0} & \bM_{\rm vib}
\end{array}
\right).
\eea
Calculating the inverse of both sides of \Eq{acmatinrs} leads to
\be
\begin{array}{l}
\bM^{-1}=\left(
\begin{array}{cc}
\bG_{\rm rot} & \bG_{\rm Cor}\\
\bG_{\rm Cor}^\dagger & \bG_{\rm vib}
\end{array}
\right)
=\left(
\begin{array}{cc}
\bM_{\rm rot} & \bM_{\rm Cor}\\
\bM_{\rm Cor}^\dagger & \bM_{\rm vib}
\end{array}
\right)^{-1}
\\
=
\left(
\begin{array}{cc}
\textbf{1} & \textbf{0}\\
-\bM_{\rm vib}^{-1}\bM_{\rm Cor}^\dagger & \textbf{1}
\end{array}
\right)
\left(
\begin{array}{cc}
\bM_{\rm rot}-\bM_{\rm Cor}\bM_{\rm vib}^{-1}\bM_{\rm Cor}^\dagger & \textbf{0}\\
\textbf{0} & \bM_{\rm vib}
\end{array}
\right)^{-1}
\left(
\begin{array}{cc}
\textbf{1} & -\bM_{\rm Cor}\bM_{\rm vib}^{-1}\\
\textbf{0} & \textbf{1}
\end{array}
\right).\\
\end{array} 
\ee
Calculating the matrix product in above equation leads to the final results
\bea\label{Minvs}
\bG_{\rm rot}&=&\left(\bM_{\rm rot}-\bM_{\rm Cor}\bM_{\rm vib}^{-1}\bM_{\rm Cor}^\dagger\right)^{-1} \nl
\bG_{\rm Cor}&=&-\bG_{\rm rot}\bM_{\rm Cor}\bM_{\rm vib}^{-1} \nl
\bG_{\rm vib}&=&\bM_{\rm vib}^{-1}+\bM_{\rm vib}^{-1}\bM_{\rm Cor}^\dagger\bG_{\rm Cor} .
\eea
The  matrix $\bG$ is essentially the same as the one reported in Ref. \cite{gatti01:8275}. There, however, the essential point has been to express the MRF Cartesian components of the angular momentum associated with each vector $\{ {\bR}_i\}$ in terms of $\bfJ$ and $\{P_{\theta_{i}},P_{\varphi_{i}}\}$.

For the special case of Jacobi vectors, $\bM_{\rm vib}$ is diagonal which greatly simplifies the calculation of \Eq{Minvs}. 
It is straightforward to obtain the diagonal elements according to \Eq{acvelocity}
\bea
&&\left(\bM_{\rm vib}\right)_{R_jR_j}=\mu_j \nl
&&\left(\bM_{\rm vib}\right)_{\theta_j\theta_j}=\mu_j R_j^2 \nl
&&\left(\bM_{\rm vib}\right)_{\varphi_j\varphi_j}=\mu_j R_j^2 \sin^2\theta_j.
\eea
The inverse matrix of $\bM_{\rm vib}$ can be obtained quite easily. We only need to calculate the inverse of a matrix of dimension 3$\times$3 as well as some matrix products to get the final results. Here we give the matrix elements for $\bG_{\rm vib}$ (Hermitian) which will be used in Section \ref{sec:app}:
\bea\label{kevib}
G_{{R_i}{R_j}}&=&\frac{\delta_{ij}}{\mu_j}, \hspace{10mm} G_{{R_i}{\theta_j}}=0, \hspace{10mm} G_{{R_i}{\varphi_j}}=0, \nl
G_{{\theta_i}{\theta_j}}&=&\frac{\delta_{ij}}{\mu_jR_j^2}+\frac{\cos(\varphi_i-\varphi_j)}{\mu_N R_N^2}, \nl
G_{{\theta_i}{\varphi_j}}&=&\frac{\cot\theta_j\sin(\varphi_i-\varphi_j)-\cot\theta_1\sin\varphi_i}{\mu_N R_N^2}, \nl
G_{{\varphi_i}{\varphi_j}}&=&\frac{\delta_{ij}}{\mu_jR_j^2\sin^2\theta_j}+\frac{1}{\mu_1 R_1^2 \sin^2\theta_1} \nl
&+&\frac{\cot\theta_i\cot\theta_j\cos(\varphi_i-\varphi_j)+\cot^2\theta_1-\cot\theta_1(\cos\varphi_i\cot\theta_i+\cos\varphi_j\cot\theta_j)}{\mu_N R_N^2}.
\eea

\begin{acknowledgments}
We gratefully acknowledge financial support by the Deutsche Forschungsgemeinschaft through the GK 788.
\end{acknowledgments}

\newpage 

\clearpage\newpage
\section*{Figure Captions}
\begin{figure}[h]
\caption{The four Jacobi vectors of the H$_3$O$_2^-$ anion which are used to define the 9D KEO.\label{fig:h3o2-coords}}
\end{figure}
\begin{figure}[h]
\caption{Definition of active coordinates of the reduced
N$_2$H$_7^+$ model.\label{fig:n2h7-coords}}
\end{figure}
\clearpage\newpage
\vspace*{5cm}
\begin{figure}[h]
\centering
\includegraphics [scale=0.35] {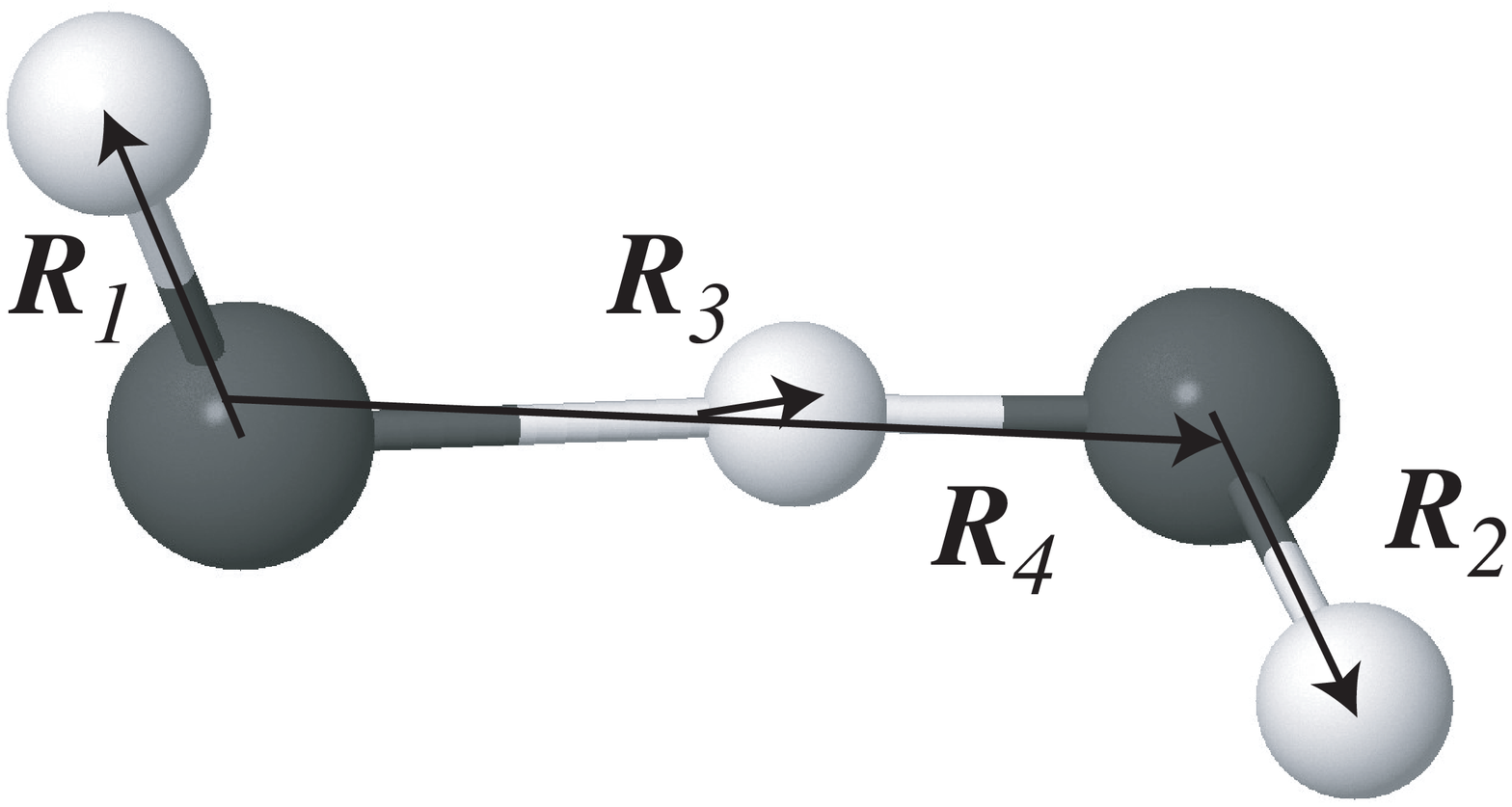}
\end{figure}
\vspace*{5cm}
Figure 1, Yang and K\"uhn
\clearpage\newpage
\vspace*{5cm}
\begin{figure}[h]
\centering
\includegraphics [scale=0.8] {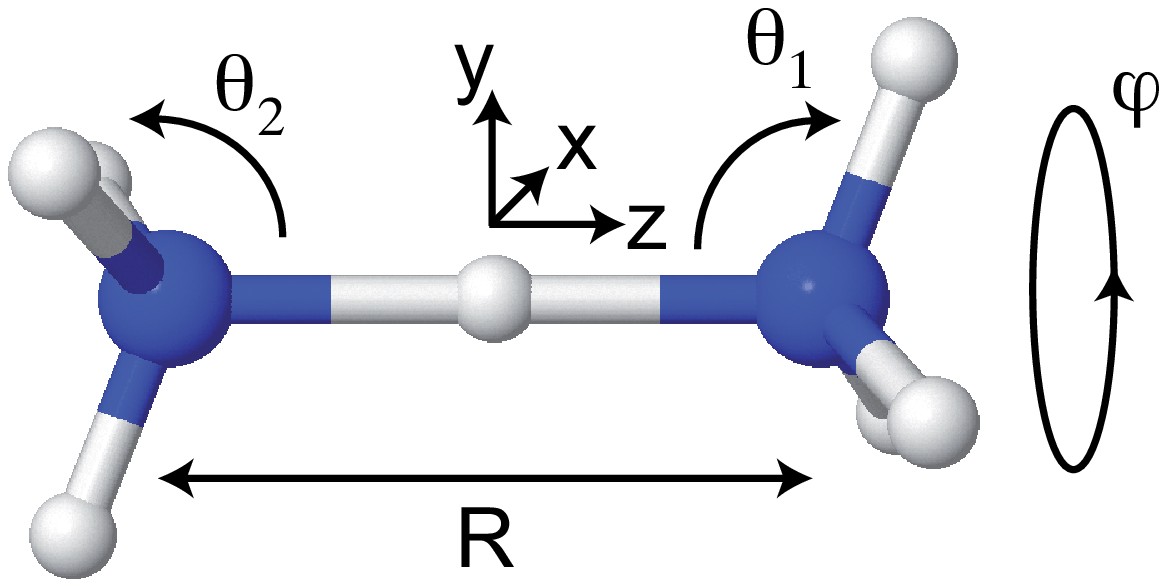}
\end{figure}
\vspace*{5cm}
Figure 2, Yang and K\"uhn

\end{document}